\documentclass[pre,aps,12pt]{revtex4}
\usepackage{graphicx}

\begin{document}

\title{Heisenberg-Ising delta-chain with bond alternation.}
\author{D.~V.~Dmitriev}
\author{V.~Ya.~Krivnov}
\email{krivnov@deom.chph.ras.ru} \affiliation{Institute of
Biochemical Physics of RAS, Kosygin str. 4, 119334, Moscow,
Russia.}
\date{}

\begin{abstract}
The spin-$\frac{1}{2}$ delta-chain (sawtooth chain) with
antiferromagnetic Heisenberg basal chain and Ising apical-basal
interactions is studied. The basal-apical interactions involve the
bond alternation. The limiting cases of the model include the
symmetrical delta-chain and the antiferromagnetic chain in the
staggered magnetic field. We study ground state properties of the
model by the exact diagonalization and density matrix
renormalization group methods. The ground state phase diagram as a
function of the bond alternation consists of magnetic and various
non-magnetic phases. All phases excluding the ferrimagnetic phase
are gapped and an origin of the gaps is cleared.
\end{abstract}

\maketitle

\section{Introduction}

\begin{figure}[tbp]
\includegraphics[width=5in,angle=0]{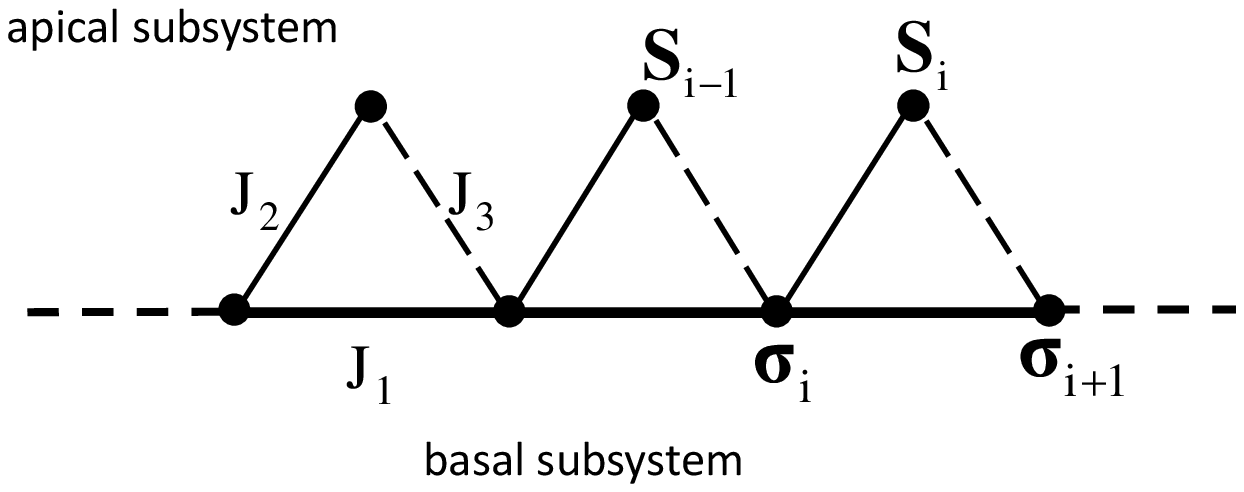}
\caption{The delta-chain model.}
\label{Fig_saw}
\end{figure}

The low-dimensional quantum magnets on geometrically frustrated
lattices attract much interest last years \cite{diep}. An
important class of such systems includes lattices consisting of
triangles. An interesting and a typical example of these systems
is the $s=\frac{1}{2}$ delta-chain consisting of a linear chain of
triangles as shown in Fig.\ref{Fig_saw}. The interaction $J_{1}$
acts between neighboring basal spins, $J_{2}$ and $J_{3}$ are
interactions of the basal ($\sigma_i$) and apical ($S_{i}$) spins.
The antiferromagnetic (AF) Heisenberg delta-chain has been studied
extensively and it demonstrates a variety of peculiar properties
\cite{sen,Nakamura,Blundell,flat,Schulen,Mak,Zhit,Schmidt,Honecker,Derzhko}.
In contrast to the AF delta-chain the same model with AF
basal-basal and ferromagnetic (F) apical-basal interactions (F-AF
delta-chain) is less studied, especially as a function of the
ratio of the F and AF interactions. An additional motivation of
the study of this model is the existence of real compounds,
malonate-bridged copper complexes \cite{Inagaki}, which are
described by this model. The F-AF delta-chain can be extended to a
model including bond alternation, when $J_{2}\neq J_{3}$. The
competition between frustration and bond alternation is of another
physical interest.

The Hamiltonian of the model has a form
\begin{equation}
\hat{H}=\sum_{i=1}^{N}\mathbf{\sigma}_{i}\cdot\mathbf{\sigma}_{i+1}
-g\sum_{i=1}^{N}\mathbf{\sigma}_{i}\cdot(\mathbf{S}_{i} +\gamma
\mathbf{S}_{i-1})  \label{isotr}
\end{equation}%
where we use the parametrization $J_{1}=1$, $J_{2}=-g$ and
$J_{3}=-g\gamma $ with $g>0$ and $0\leq \gamma \leq 1$.

By now not much is known about the ground state properties of this
model. The studies of model (\ref{isotr}) without the bond
alternation ($\gamma =1$) \cite{Tonegawa} show that the ground
state is ferromagntic for $g>2$. It is supposed \cite{Tonegawa}
that it is ferrimagnetic for $g<2$, though there is no analytical
proof and numerical calculations do not give reliable prediction
due to strong finite size effects. The critical point $g=2$
separates two phases and the ground state in this point is
macroscopically degenerate \cite{degeneracy}.

The model with bond alternation on delta chain (\ref{isotr}) was
not studied before. It can be shown that the ferromagnetic ground
state is stable for the model with bond alternation (\ref{isotr})
in the region $g>g_{0}(\gamma)$ with $g_{0}=1+\frac{1}{\gamma }$.
That is the critical point $g=2$ extends into the transition line
$g_{0}(\gamma)$ (with the same total number of degenerate states).
The analytical study of the ground state properties of the model
(\ref{isotr}) for $0<\gamma <1$ and $g<g_{0}$ is very complicated
problem and numerical calculations meet similar problems as for
the symmetric $\gamma =1$ case. From this point of view it is
useful to consider more simple model preserving main qualitative
features of the initial model. The simplification consists in the
replacement of some Heisenberg interactions with the Ising terms.
We take such replacement for basal-apical interactions. As a
result, model (\ref{isotr}) reduces to the spin-$\frac{1}{2}$
delta chain with the antiferromagnetic (AF) Heisenberg basal-basal
interaction and the ferromagnetic Ising basal-apical interactions
(Fig.\ref{Fig_saw}) with bond alternation. The Hamiltonian of the
model has a form
\begin{equation}
\hat{H}=\sum_{i=1}^{N}\mathbf{\sigma}_{i}\cdot\mathbf{\sigma}_{i+1}
-g\sum_{i=1}^{N}\sigma _{i}^{z}(S_{i}^{z}+\gamma S_{i-1}^{z})
\label{H}
\end{equation}

In this paper, we report results of our numerical calculations of
the ground state and the low energy excitations of Hamiltonian
(\ref{H}) using the exact diagonalization (ED) of finite chains and
the density matrix renormalization group (DMRG) method.

The paper is organized as follows. In Section II we reduce model
(\ref{H}) to AF chain in non-uniform magnetic field. Two special
cases of model (\ref{H}): symmetric case $\gamma=1$ and the case
$\gamma=0$ are studied in Section III and Section IV,
respectively. In Section V the ground state phase diagram of model
(\ref{H}) is constructed. In Section VI we give a summary of our
results.

\section{Ising interaction as magnetic field.}

According to Eq.(\ref{H}) the considered model is the
antiferromagnetic (AF) basal spin-$\frac{1}{2}$ chain in an
external non-homogenous magnetic field induced by apical spins.
The local magnetic field $h_{i}$ acting on $i$-th basal spin is
$h_{i}=g(S_{i}^{z}+\gamma S_{i-1}^{z})$. Depending on the
configuration of two adjacent apical spins the local magnetic
field can be $h_{i}=\pm h_{uf}$ or $h_{i}=\pm h_{st}$ where
\begin{eqnarray}
h_{uf} &=&\frac{(1+\gamma )g}{2}  \label{h} \\
h_{st} &=&\frac{(1-\gamma )g}{2}  \nonumber
\end{eqnarray}

Generally, in order to find the ground state of model (\ref{H})
one needs to go through the following three steps: 1) to take
definite configuration of the apical spin subsystem, 2) then to
calculate the ground state energy of the AF chain in the induced
magnetic field and, finally, 3) to choose the spin configuration
of the apical spin subsystem which provides the lowest energy. We
name such configuration as an optimal one. Obviously, it is
impossible to examine all $2^{N}$ configurations of the apical
spins. Therefore, we need to choose and study the most important
classes of the apical spin configurations. Our numerical
calculations indicate that the optimal apical configuration
depends on both the basal-apical interaction $g$ and the
bond-alternation parameter $\gamma $. But for any parameters $g$
and $\gamma $ it is either ferromagnetic one or it belongs to the
class of periodic arrangement of the up and down apical spins.

The ferromagnetic apical spin configuration with
$S^{z}=\frac{N}{2}$ ($S^{z}=\sum_{i=1}^{N}S_{i}^{z}$) reduces the
model (\ref{H}) to the AF basal spin chain in the uniform magnetic
field $h_{uf}$. This model is exactly solvable one and has
non-zero magnetization $\sigma ^{z}=\sum_{i=1}^{N}\sigma _{i}^{z}$
depending on the value of $h_{uf}$. So that the magnetization of
the total system is $L^{z}=\frac{N}{2}+\sigma^{z}$.

The periodic apical configurations are obviously non-magnetic ones
($S^{z}=0$) and can not induce magnetization of the basal
subsystem. Therefore, the periodic apical configurations produce
non-magnetic states $L^{z}=0$ independent of the period length.

As a function of $\gamma $ model (\ref{H}) interpolates
between the symmetric delta-chain at $\gamma =1$
and the AF chain in a staggered field at $\gamma =0$. The
properties of the model are essentially different in these
limiting cases. Therefore, we study these special cases separately
before the constructing of the ground state phase diagram of model
(\ref{H}) in the general case.

\section{Symmetric delta-chain, $\gamma =1$}

When $\gamma =1$ the uniform field is $h_{uf}=g$, while the
staggered field vanishes $h_{st}=0$. At first we consider the
ferromagnetic apical spin configuration. As noted above, in
this case model
(\ref{H}) reduces to the AF chain in an uniform magnetic field
$h_{i}=g$. The energy of the state with the basal spin $\sigma
^{z}=N\sigma $ and the total spin $L^{z}=$ $N(\sigma
+\frac{1}{2})$ is
\begin{equation}
E(\sigma )=E_{0}(\sigma )-Ng\sigma
\end{equation}%
where $E_{0}(\sigma )$ is the energy of the ground state of the AF
chain in the spin sector $\sigma ^{z}=N\sigma $. The lowest energy
of the AF chain in the uniform field $h_{uf}=g$ is reached for
$\sigma =\sigma _{0}$ and the value $\sigma _{0}(g)$ is determined
by the condition
\begin{equation}
\frac{dE_{0}(\sigma_0)}{d\sigma}=N g
\end{equation}%
and can be found from the solution of the Bethe-ansatz equations. The
energy $E(\sigma )$ of the state in the uniform field $h_{uf}=g$
with $\sigma =\sigma _{0}$ we denote as $E_{uf}$. The total spin
of this state is $L_{0}^{z}=N(\frac{1}{2}+\sigma _{0})$.

Model (\ref{H}) with the periodic configurations of the apical
spins is the AF chain in a modulated magnetic field $h(q)$
\begin{equation}
h(q)=\frac{1}{N}\sum_{n=1}^{N}\exp (-iqn)h_{n}
\end{equation}%
with a period $l=2\pi /q$. The exact solution of the AF Heisenberg
chain in a modulated magnetic field is unknown generally and
therefore we employ numerical calculations. These calculations show
that in the periodical magnetic field with $S^{z}=0$ the spin of
the lowest state of (\ref{H}) is $L^{z}=0$. Therefore, we wait
that the ground state spin of model (\ref{H}) at given $g$ can be
either $L^{z}=0$ or $L_{0}^{z}=N(\frac{1}{2}+\sigma _{0}(g))$.

Let us consider the dependence of the energy on the periodicity of
the field, $E(l)$. If $l=2$ $(q=\pi )$ then $h_{n}=(-1)^{n}h_{st}$
and the modulated field is a staggered one. But for $\gamma =1$
$h_{st}=0$ and the staggered field is not effective in this case.
If $l=4$ $(q=\frac{\pi}{2})$ the configuration of the apical spins
is $\uparrow \uparrow \downarrow \downarrow \uparrow \uparrow
\ldots $ and $h_{n}=g\cos (\frac{\pi n}{2})$. For $g\ll 1$ this
problem can be solved if a zero-field susceptibility $\chi (q)$ is
known. The energy of AF chain in the magnetic field $h(q)$ is
\begin{equation}
E(l)=E_{0}(0)-\frac{N\chi (q)h^{2}(q)}{2}  \label{Energy}
\end{equation}
where $E_{0}(0)=-N(\ln 2-\frac{1}{4})$ and $q=2\pi /l$.

The exact expression for $\chi (q)$ is unknown and the
approximation for $\chi (q)$ is proposed in \cite{Muller} which is
\begin{equation}
\chi (q)=\frac{q}{\pi ^{2}\sin q}  \label{chi}
\end{equation}

\begin{figure}[tbp]
\includegraphics[width=4in,angle=-90]{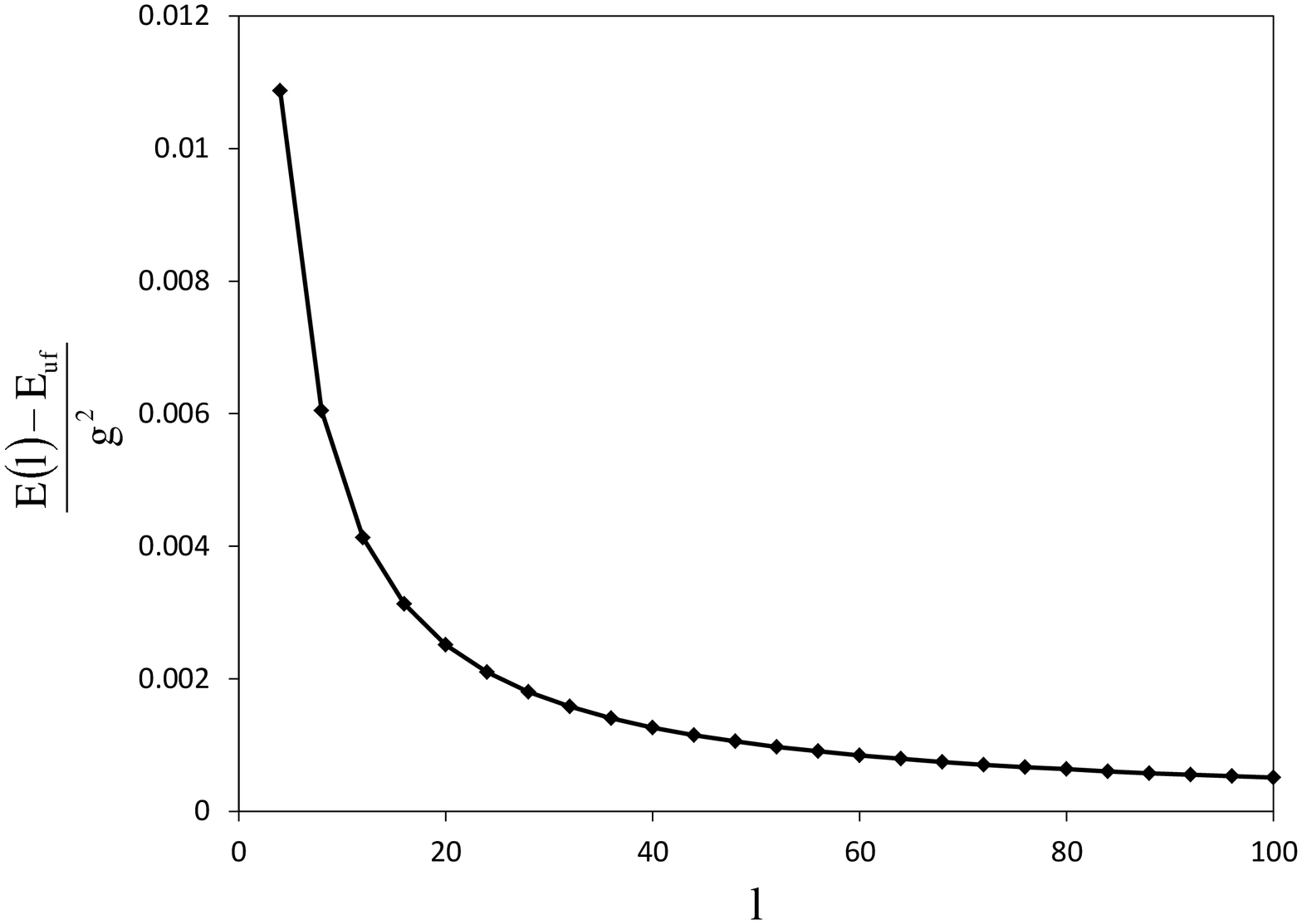}
\caption{Energy of AF basal chain in the modulated magnetic field
with period $l$ induced by the apical spins. $E_{uf}$ is the
energy in the uniform field $h_{uf}=g$.} \label{Fig_ELh0}
\end{figure}

\begin{figure}[tbp]
\includegraphics[width=4in,angle=-90]{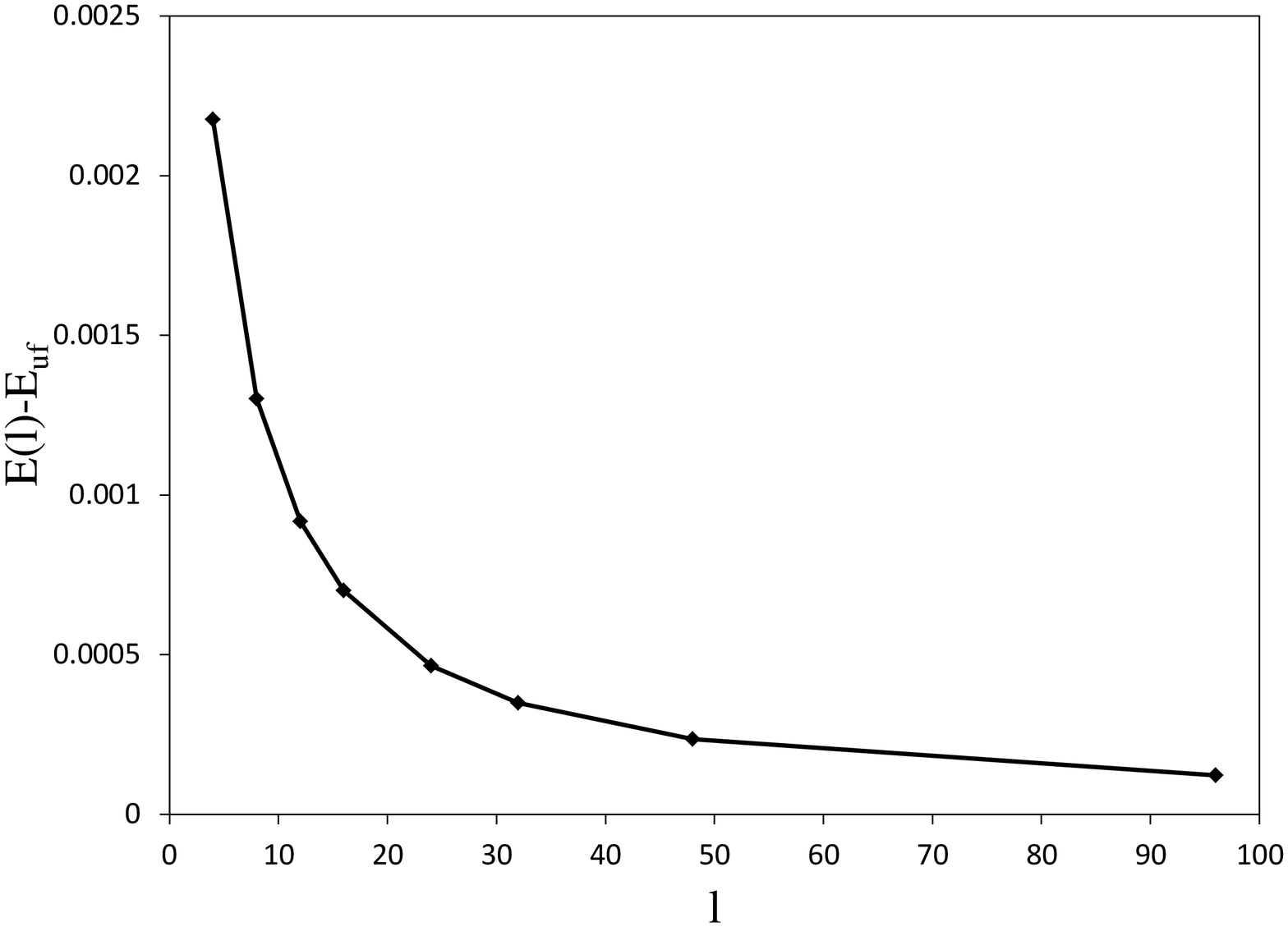}
\caption{Energy of model (\ref{H}) for periodic configurations of
apical spins with period $l$. Numerical data are obtained for
$g=0.5$ by DMRG calculations for $N=96$.} \label{Fig_ELh05}
\end{figure}

Eq.(\ref{chi}) gives qualitatively correct dependence of
$\chi(q)$. Energy (\ref{Energy}) as a function of the periodicity
of the modulated field calculated with the use of Eq.(\ref{chi})
at $g\ll 1$ is shown in Fig.\ref{Fig_ELh0}. Similar dependence for
$g=0.5$ obtained by numerical calculations is presented in
Fig.\ref{Fig_ELh05}. These results show that the energy decreases
when the period $l$ increases. The minimal energy is reached for
the configuration with $\frac{N}{2}$ apical spins up and
$\frac{N}{2}$ spins down $\uparrow \uparrow \uparrow \ldots
\uparrow \downarrow \downarrow \downarrow \ldots \downarrow $
(two-domain structure). Formally, this configuration corresponds
to the period $l=N$. The magnetic field $h_{i}$ induced by this
spin configurations in the cyclic basal chain is
\begin{eqnarray}
h_{1} &=&h_{N/2}=0  \label{2-domain} \\
h_{n} &=&g,\quad 1<n<\frac{N}{2}  \nonumber \\
h_{n} &=&-g,\quad \frac{N}{2}<n<N  \nonumber
\end{eqnarray}

There are $N$ such states and their total spin is $L^{z}=0$.

\begin{figure}[tbp]
\includegraphics[width=4in,angle=0]{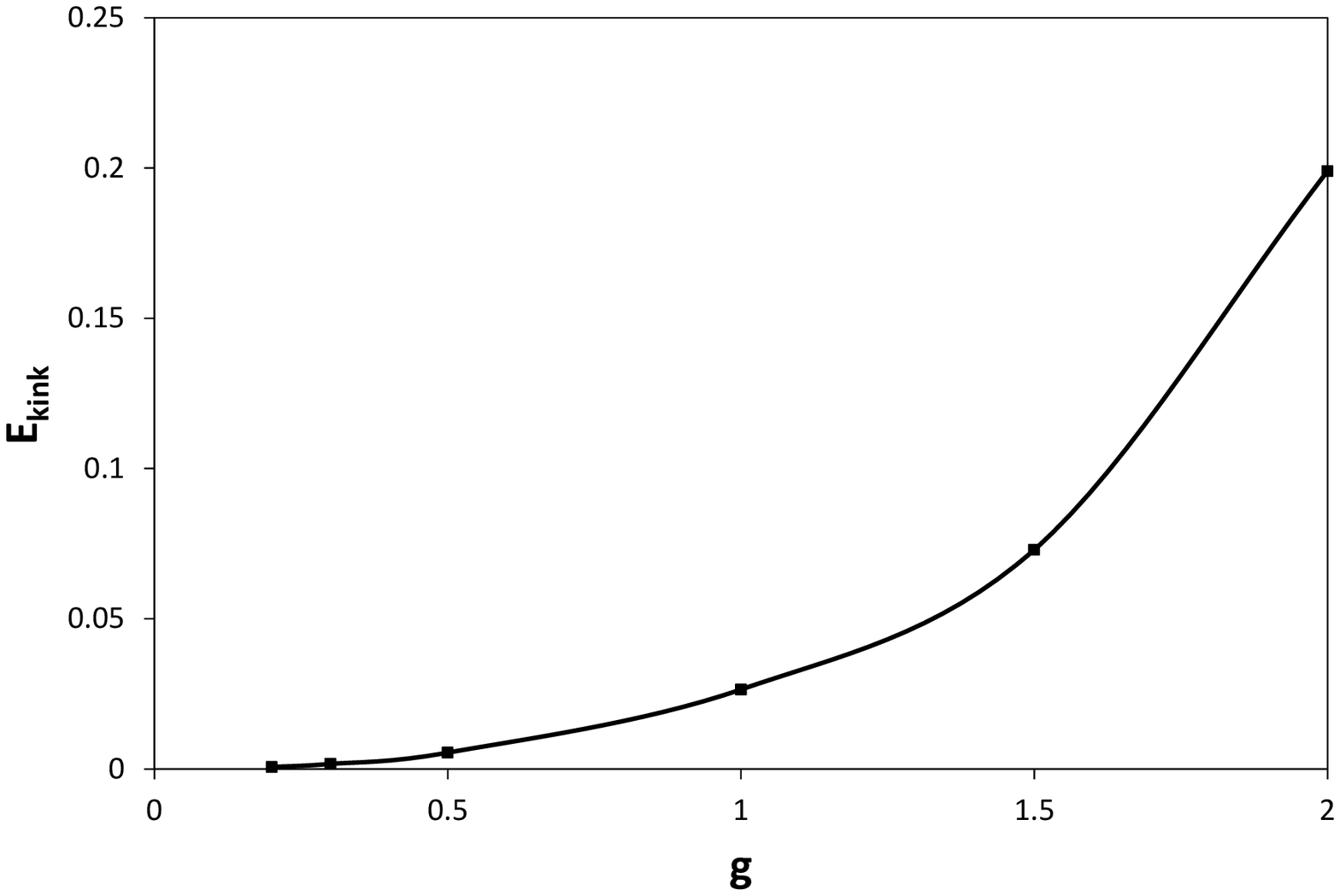}
\caption{Kink energy as a function of $g$ for the symmetric case
$\gamma=1$.} \label{Fig_Ekink}
\end{figure}

As it is seen in Fig.\ref{Fig_ELh05} the energy in the field
(\ref{2-domain}) $E_{l=N}$ is higher than the energy $E_{uf}$.
Really the latter is the lowest energy at given $g$ and the ground
state spin is $L_{0}^{z}(g)=N(\frac{1}{2}+\sigma _{0}(g))$.
However, the energy of the state with $L^{z}=0$ in the field
(\ref{2-domain}) approaches to that in the uniform field at
$N\to\infty $. It is not surprising because the local
magnetization $\left\langle \sigma _{n}^{z}\right\rangle $ in one
half of the basal system is $\sigma _{0}$ and in another one is
$(-\sigma _{0})$ at $N\gg 1$. But the energy of this state is
still higher than the ground state energy due to the presence of
two domain walls (`kinks'), where the magnetic field
(\ref{2-domain}) is zero. The kink energy $E_{kink}$ is a half of
the difference between the energy $E_{l=N}$ in the field
(\ref{2-domain}) and the energy $E_{uf}$ in the uniform field
$h_{uf}=g$. Therefore, $E_{l=N}=E_{uf}+2E_{kink}$. The dependence
of the kink energy $E_{kink}$ on $g$ obtained by the numerical
calculations is shown in Fig.\ref{Fig_Ekink} for $g<2$. As follows
from Fig.\ref{Fig_Ekink} the kink energy tends to zero when $g\to
0$, because in this limit model (\ref{H}) reduces to the gapless
AF basal chain.

\begin{figure}[tbp]
\includegraphics[width=4in,angle=0]{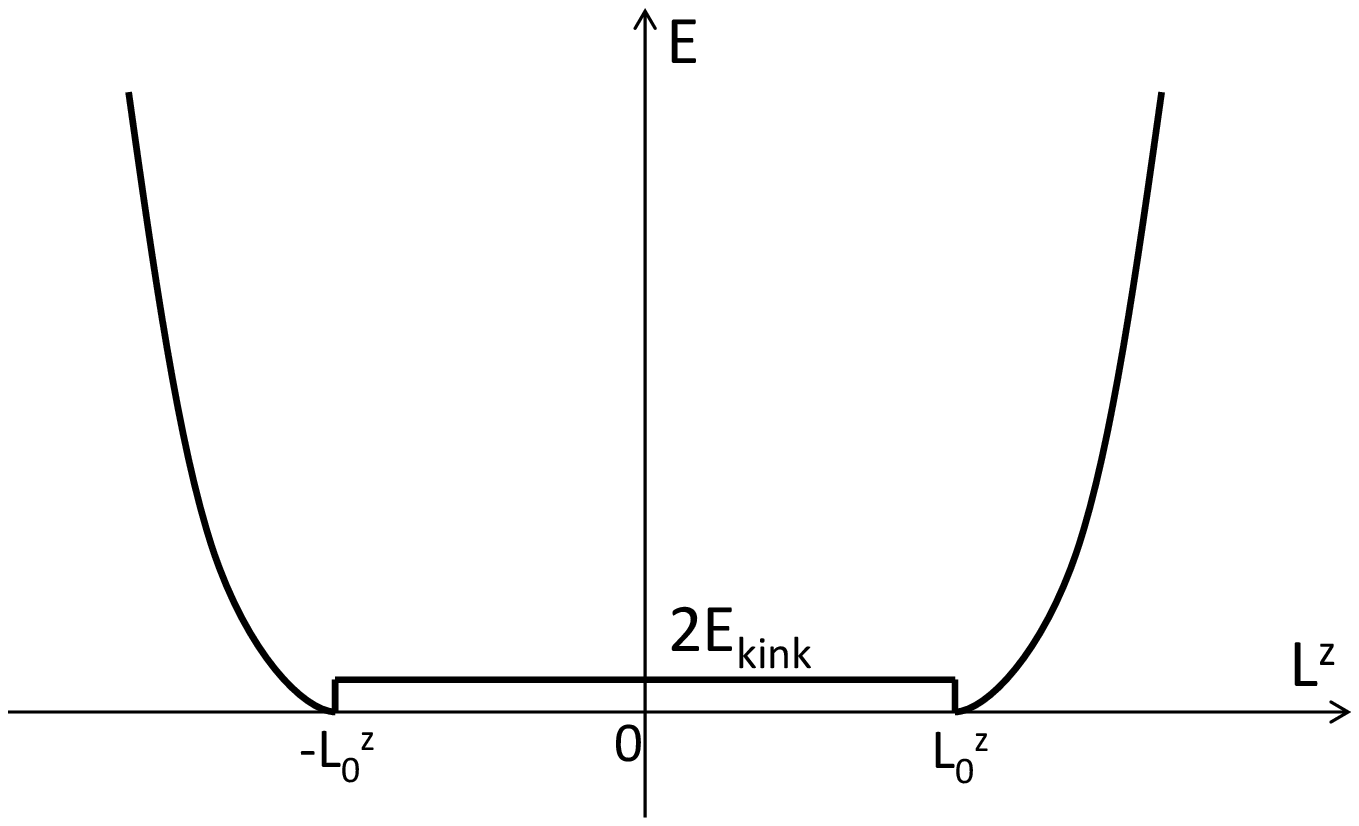}
\caption{The schematic spectrum $E(L^z)$ of the symmetric model
($\gamma=1$) in the ferrimagnetic phase.} \label{Fig_ESz}
\end{figure}

Similarly, one can construct the two-domain configuration with $k$
apical spins down and $(N-k)$ up. If $k\gg 1$ then the energy of
this state is $E_{k}=E_{N/2}+O(N^{-1})$ and the total spin is
$L^{z}=(1-2k/N)L_{0}^{z}(g)$, i.e. $\left\vert L^{z}\right\vert
<L_{0}^{z}(g)$. It can be shown \cite{anisotrop} that such
two-domain configurations are the ground state in the spin sectors
$\left\vert L^{z}\right\vert <L_{0}^{z}(g)$. In the spin sector
$\left\vert L^{z}\right\vert >L_{0}^{z}(g)$ the ground state is
reached by the ferromagnetic apical configuration and has the
energy $(E_{0}(\sigma )-Ng\sigma)$. But the global ground state of
model (\ref{H}) is two-fold degenerate with total spin $\pm
L_{0}^{z}(g)$ and energy $E_{uf}$. The ground state energy for
$\gamma =1$ as a function of the total spin $L^{z}$ at fixed value
$g<2$ is shown schematically in Fig.\ref{Fig_ESz}.

\section{The case $\gamma =0$}

For $\gamma =0$ the fields $h_{uf}$ and $h_{st}$ are
$h_{uf}=h_{st}=\frac{1}{2}g$. In this limiting case the spins are
located on a bipartite lattice with equal number of spins on
sublattices. According to the Lieb-Mattis theorem
\cite{Lieb,Ovchinnikov} the ground state of model (\ref{H}) is in
the $L^{z}=0$ spin sector. Strictly speaking this theorem is
applicable to the models in which spin-spin interactions contain
transverse components rather than only the Ising ones. However,
even an infinitesimal transverse basal-apical interaction $J_{xy}$
leads to the $L^{z}=0$ ground state. Because this state cannot be
destroyed by the infinitesimal perturbation it remains as the
ground state in the spin sector $L^{z}=0$ for $J_{xy}=0$. The
numerical calculations confirm this statement. The only difference
between $J_{xy}=0$ and $J_{xy}\neq 0$ cases is related to the
degeneracy of the $L^{z}=0$ ground state. When $J_{xy}\neq 0$ it
is non-degenerate but for $J_{xy}=0$ it is two-fold degenerate:
the configuration of the apical spins corresponds to the staggered
field with $h_{n}=\pm \frac{(-1)^{n}}{2}g$ and such configuration
is optimal for all values of $g$. For example, the ground state
energy $E_{st}$ in the staggered field for $g\ll 1$ is
proportional to $-g^{4/3}$ rather than to $-g^{2}$ as for the
model in the uniform or periodic field with $l>2$. For $g\gg 1$
the energy as a function of the periodicity of the field $l$ can
be calculated analytically:
\begin{equation}
E(l)=-N\frac{g-1}{4}-\frac{N}{l}(1+\frac{1}{2g})
\end{equation}
and it is minimal for the staggered configuration ($l=2$).

\section{The ground state phase diagram for $0<\gamma <1$}

As follows from the above the optimal apical configuration and
corresponding magnetic field acting on the AF chain depends on
bond alternation. For $\gamma =1$ this field is uniform and the
ground state is ferro- or ferrimagnetic depending on whether $g>2$
or $g<2$ and the total ground state spin $L^{z}$ $\neq 0$. But for
$\gamma =0$ the optimal magnetic field is staggered and $L^{z}=0$.
Therefore, the transition between magnetic ($L^{z}\neq 0$) and
non-magnetic ($L^{z}=0$) ground state phases occur somewhere in
the region $0<\gamma <1$.

In order to obtain a qualitative representation of the phase
diagram in ($\gamma,g$) plane at first we consider the XX variant
of model (\ref{H}) with the Hamiltonian
\begin{equation}
\hat{H}=\sum_{i=1}^{N}(\sigma _{i}^{x}\sigma _{i+1}^{x}+\sigma
_{i}^{y}\sigma _{i+1}^{y})-g\sum_{i=1}^{N}\sigma
_{i}^{z}(S_{i}^{z}+\gamma S_{i-1}^{z}) \label{HXX}
\end{equation}

\begin{figure}[tbp]
\includegraphics[width=5in,angle=0]{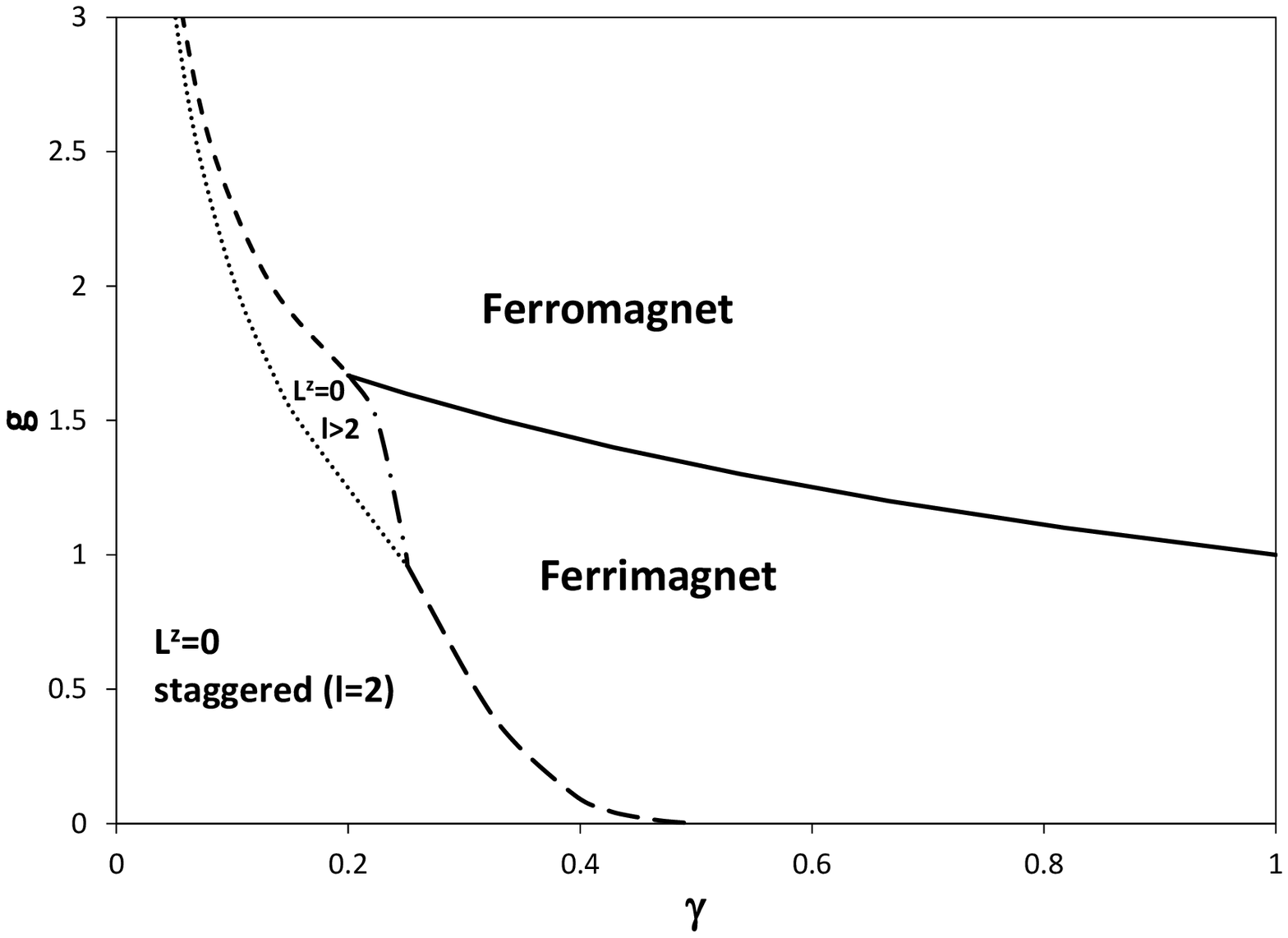}
\caption{Phase diagram of the spin $\Delta$-chain with $XX$
interaction on basal chain and Ising interaction between basal and
apical chains.} \label{Fig_phaseXX}
\end{figure}

The $XX$ model in the periodic magnetic field with period $l$ can
be diagonalized through the Jordan-Wigner transformation and by
the construction of $l$ reduced Brilluene zones. We omit technical
details and represent the ground state phase diagram of
Eq.(\ref{HXX}) in $(\gamma,g)$-plane in Fig.\ref{Fig_phaseXX}. The
ground state phase diagram consists of the ferromagnetic,
ferrimagnetic phases and the non-magnetic phase with optimal
staggered apical field. The phases are divided by the intermediate
regions or boundary lines. For $0.5<\gamma <1$ the ground state is
ferromagnetic or ferrimagnetic and the boundary between these
magnetic phases lies on $g=\frac{2}{1+\gamma }$ line. The
transitions between the magnetic and non magnetic phases occur at
$0<\gamma <0.5$.

\begin{figure}[tbp]
\includegraphics[width=4in,angle=-90]{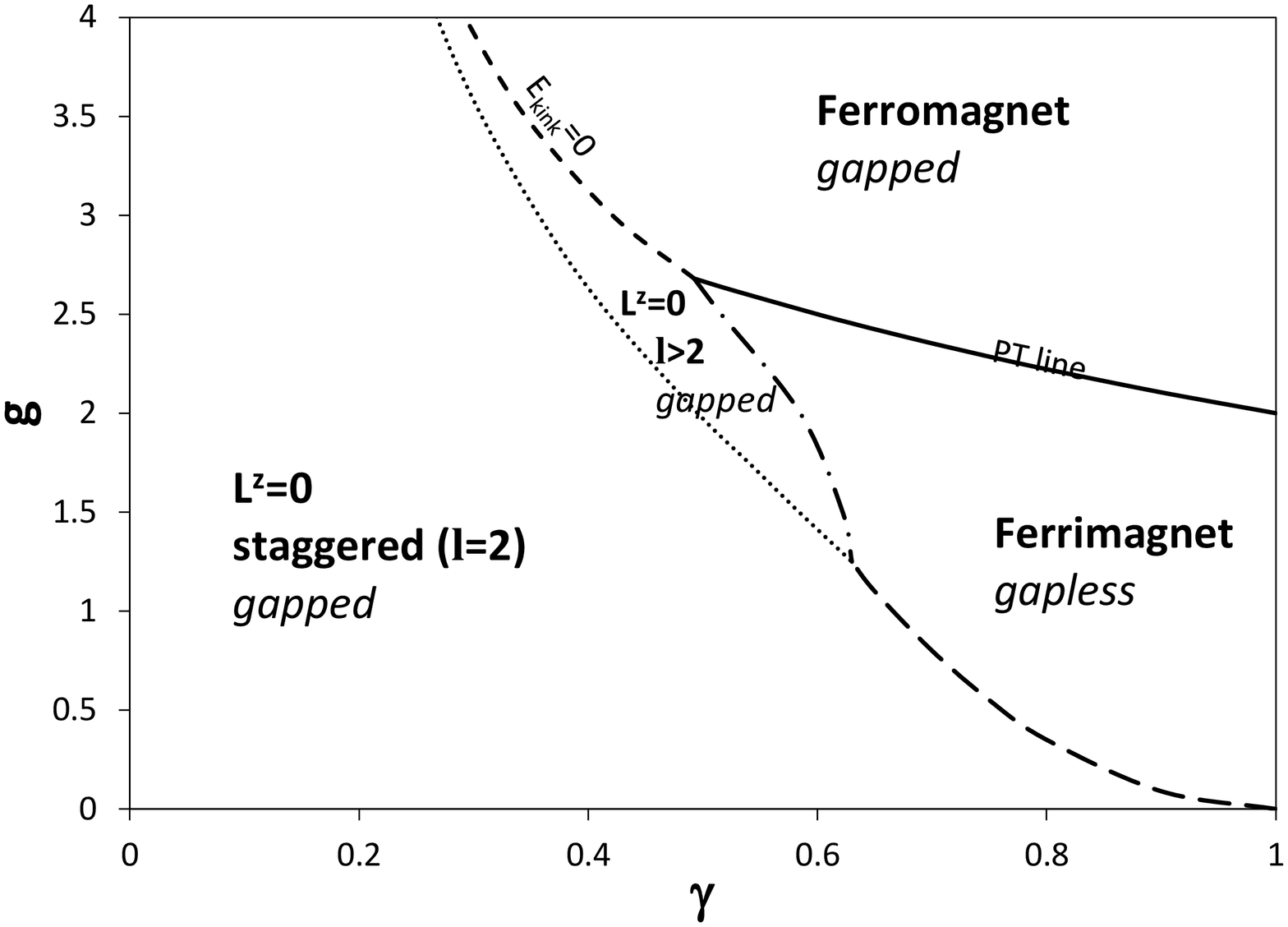}
\caption{The phase diagram of the spin $\Delta$-chain with AF
Heisenberg interactions on basal chain and Ising interaction
between basal and apical chains. Dotted line is the boundary
between the staggered and $l=4$ phases. Dashed line separates
ferro- and non-magnetic phases, the kink energy is zero on this
line. Long dashed line is the boundary between ferrimagnetic and
staggered phases. In the intermediate region the phases with $l>2$
exist. The interfacial lines are obtained by extrapolation of
numerical data of ED (up to N=24) and DMRG (up to N=96).}
\label{Fig_phase}
\end{figure}

The phase diagram of Eq.(\ref{H}) (the AF chain in the apical
field) is obtained on a base of numerical calculations and it is
shown in Fig.\ref{Fig_phase}. The phase diagram of this model is
qualitatively very similar to that for the XX model. Besides, the
main features of both phase diagrams are the same except shifts of
the boundary lines and intermediate regions. For example, the
region of the existence of the non-magnetic phase for the XX model
is shifted from $0<\gamma <1$ to $0<\gamma <0.5$ in comparison
with that for the AF model.

Now we discuss the ground state phase diagrams of the AF model
shown in Fig.\ref{Fig_phase}. For brevity we will refer the ground
state phase in the optimal periodical field with period $l$ as the
$l$-phase. For example, if the optimal field is staggered the
ground state phase is the staggered one or the ($l=2$) phase. The
ground state phases for which the optimal field is uniform are
ferromagnetic (fully polarized) or ferrimagnetic (partly
polarized) ones. This phase diagram consists of the ferromagnetic,
the ferrimagnetic and the staggered phases as well as the
intermediate regions between them. The ferromagnetic and the
ferrimagnetic phases are separated by the critical line of the
Pokrovsky-Talapov type (PT-line). On this line
$g=\frac{4}{1+\gamma }$ and $h_{uf}=2$. The ferrimagnetic phase is
gapless up to the PT line and the ferromagnetic phase is gapped.
The first excited state in the F phase is in the total spin sector
$L_{z}=N-1$ as long as $2E_{kink}>(h_{uf}-2)$ (the kink in the
ferromagnetic phase is defined exactly as for the ferrimagnetic
phase) or in the sector $L_{z}=0$ if $2E_{kink}<(h_{uf}-2)$. The
gap between the first excited state and the ground state is
$\Delta E=(h_{uf}-2)$ or $\Delta E=2E_{kink}$, respectively. The
line where $E_{kink}=0$ (dashed line in Fig.\ref{Fig_phase})
intersects the PT line and these two lines form the boundary of
the ferromagnetic phase.

The boundaries of the staggered phase are shown by dotted and long
dashed lines. The dotted line is the transition between the
staggered ($l=2$) and ($l=4$) phases whereas long dashed line
separates the staggered and the ferrimagnetic phases.

The transition from the staggered phase to the ferromagnetic phase
occurs through the intermediate region. When $\gamma $ increases
from $\gamma =0$ (at fixed $g$) the staggered phase passes to the
($l=4$) phase, then successive transitions occur from the phases
($l=4$) to ($l=6$), then from ($l=6$) to ($l=8$) and so on up to
the ferromagnetic phase. In other words, an infinite series of
phases are present in the intermediate region between the
staggered and the ferromagnetic phases. In the part of the
intermediate region between the staggered and the ferrimagnetic
phases the transitions from the ($l=4$) phase to other phases
occur up to the phase with some $l=l^{\ast }(\gamma)$ and then
from the $l^{\ast }$-phase to the ferrimagnetic phase, i.e. a
finite number of phases exist in this region. Below the
intermediate region on the long dashed line the direct transition
from the staggered to the ferrimagnetic phase takes place. The
form of this transition line at $g\ll 1$ can be estimated as
follows. For $g\ll 1$ the ground state energy $E_{uf}$ in the
ferrimagnetic phase is
\begin{equation}
E_{uf}\simeq E_{0}(0)-N\frac{g^2(1+\gamma)^2}{8\pi^2}
\end{equation}

But the ground state energy in the staggered field ($l=2$) is
proportional to $[g(1-\gamma )]^{4/3}$ and the transition between
the staggered and the ferrimagnetic phases occurs at $\gamma $
close to $1$. The value of $\gamma $ at which the ground state
energies in the uniform and the staggered fields coincide with
each other can be obtained using the known results \cite{Affleck}
about the dependence of the ground state energy in the staggered
field $E_{st}$ on $h_{st}$ at $h_{st}\ll 1$. According to
\cite{Affleck} $E_{st}-E_{0}(0)\simeq -0.29Nh_{st}^{4/3}$.
Therefore, equation $E_{st}=E_{uf}$ defines the transition line as
$g\simeq 2.7(1-\gamma )^{2}$ at $\gamma\to 1$.

The ground state of model (\ref{H}) for any values of $g$ and
$\gamma $ is realized in the periodic or the uniform
configurations of the apical spins but the problem relating to the
energy gap is more complicated. The gap is a difference between
the ground state energy and the energy of the lowest excited
state. The configuration of the apical spins for the lowest
excitation can be different from that for the ground state. This
feature has been noted before for the gap in the ferromagnetic
phase. Similar property of the gap holds in the staggered phase.
Model (\ref{H}) in the fixed staggered field is gapped
\cite{Affleck} and the first excited state is in the spin sector
$L_{z}=1$ ($\sigma _{z}=1$,$S_{z}=0$). However, a deviation of the
apical spin configuration from the ideal staggered one changes the
lowest excited state. For example, for $\gamma =0$ such deviations
are one apical spin flip ($\uparrow \downarrow \uparrow \downarrow
\uparrow \downarrow \longrightarrow \uparrow \downarrow \uparrow
\uparrow \uparrow \downarrow $) or a flip-flop of a pair of spins
($\uparrow \downarrow \uparrow \downarrow \uparrow \downarrow
\longrightarrow \uparrow \downarrow \uparrow \uparrow \downarrow
\downarrow $). An analysis of numerical results shows that the
first excited state in this case is lower than that for the ideal
staggered configuration. Such change of the first excited state is
a precursor of an instability of the staggered ground state with
respect to the breakdown of the optimal staggered field leading
finally to other ground states with the increase of $\gamma $.
Besides, the gap decreases with the increase of $\gamma$ (for
fixed $g$) and it vanishes on the interfacial lines. Similar gap
behavior occurs for all $l$- phases in which deviation from the
ideal $l$-configuration of the apical subsystem leads to a
decrease of the first excited state. Generally, all phases
excluding the ferrimagnetic phase are gapped.

\section{Summary}

We have carried out numerical calculations to study the ground
state and the energy gap in the Heisenberg-Ising delta-chain. The
replacement of the Heisenberg basal-apical interactions by the
Ising ones leads to the spin-$\frac{1}{2}$ AF Heisenberg chain in
the external magnetic field depending on the apical spin
configuration. We show that the ground state of this model has
been reached in the uniform or periodic apical spin configuration.
However, the energy gap can be induced by the apical spin
configurations different from those leading to the ground state.
We study the bond-alternation effect on the ground state phase
diagram. For the symmetric Heisenberg-Ising delta-chain the ground
state is ferro- or ferrimagnetic. The bond alternation leads to a
cascade of the phase transitions between the uniform and the
staggered ground states.

Though we considered the model with the ferromagnetic basal-apical
interaction, the properties of the model with the
antiferromagnetic interactions are the same because the model is
invariant under simultaneous change of sign of the interaction and
the substitution $S_{i}^{z}\to -S_{i}^{z}$.

In this paper we studied the delta-chain with alternating
basal-apical ferromagnetic interactions, i.e. for $0\leq \gamma
\leq 1$. The consideration can be extended to the case $\gamma
\leq 0$, which corresponds to the alternation of the F and AF
interactions. The analysis of this case shows that the ground
state is realized in the staggered apical field for all values of
$\gamma$ in the interval $-1\leq \gamma \leq 0$.

We believe that many peculiar features of the considered model
with Ising type of basal-apical interaction remains for more
complicated model with the Heisenberg basal-apical interactions.

The numerical calculations were carried out with use of the ALPS
libraries \cite{alps}.

\end{document}